\documentclass[preprint,journal]{vgtc}            % preprint (journal style)

\onlineid{2099}

\usepackage{xspace} 
\usepackage{graphicx}
\usepackage{caption,subcaption}
\usepackage{xcolor}
\usepackage{soul}
\usepackage{tikz}
\definecolor{timingColor}{HTML}{7C4085}% 984ea3
\definecolor{pairwise}{HTML}{42993F} % 4daf4a
\definecolor{errorRate}{HTML}{e41a1c}
\definecolor{usability}{HTML}{ff7f00}
\definecolor{presence}{HTML}{a65628}
\definecolor{workload}{HTML}{3172A8} % 377eb8
\newcommand{\added}[1]{{\color{black}#1}}

\definecolor{colorTodo}{RGB}{225,63,63}

\definecolor{colorWB}{RGB}{75,134,180}

\newcommand{\bl}{brushing and linking\xspace}
\newcommand{\Bl}{Brushing and linking\xspace}
\newcommand{\outline}{\textbf{O}\xspace}
\newcommand{\animated}{\textbf{A}\xspace}
\newcommand{\link}{\textbf{L}\xspace}
\newcommand{\ciicon}[1][pairwise]{%
  \begin{tikzpicture}[baseline=-0.5ex, x=1em, y=1em]
    \draw[#1, line width=0.4pt] (-0.7,0) -- (0.7,0);
    \filldraw[#1] (0,0) circle (2pt); 
  \end{tikzpicture}%
}
\vgtccategory{Research}

\vgtcpapertype{evaluation}

\title{Situated Brushing and Linking in Virtual and Augmented Reality}

\author{%
  \authororcid{Carlos Quijano-Chavez}{0000-0001-9129-5366},
  \authororcid{Benjamin Lee}{0000-0002-1171-4741},
  \authororcid{Nina Doerr}{0000-0003-3249-5354},
  \authororcid{Wolfgang Büschel}{0000-0002-3548-723X},
  \authororcid{Michael Sedlmair}{0000-0001-7048-9292}, and \\
  \authororcid{Dieter Schmalstieg}{0000-0003-2813-2235}
}

\authorfooter{
  \item
  	Carlos Quijano-Chavez, Nina Doerr, and Wolfgang Büschel are with the University of Stuttgart.
  	E-mails: quijancr@visus.uni-stuttgart.de, nina.doerr@visus.uni-stuttgart.de, wolfgang.bueschel@visus.uni-stuttgart.de.
  \item
  	Benjamin Lee is with JPMorganChase.
  	E-mail: benjammin.lee@jpmchase.com.

  \item Michael Sedlmair and Dieter Schmalstieg are with the University of Stuttgart.
  E-mails: michael.sedlmair@visus.uni-stuttgart.de, schmaldr@visus.uni-stuttgart.de.
}

\abstract{%
In traditional visual analysis, \bl is commonly used to visually connect multiple views using highlighting techniques. However, \bl has rarely been used in situated analytics, which uses visualizations to analyze data in the context of physical referents. In situated analytics, data representations must be visually linked to real-world objects. Previous work has assessed situated \bl in a virtual reality simulation of a supermarket scenario. Here, we replicate and extend the previous approach by studying \bl in an actual physical space with augmented reality, while further improving the highlighting techniques. Using a video see-through display, we compare augmented reality with virtual reality. Results suggest that AR performs better in time and accuracy, but the effectiveness of the techniques varies by condition. These results provide a new framing of how the real-world stimuli matter in situated analytics.
}


\teaser{
  \centering
  \includegraphics[width=\linewidth, alt={Snapshots showing the highlighting techniques studied in this work. The left two columns are in augmented reality, while virtual reality snapshots are in the right two columns.}]{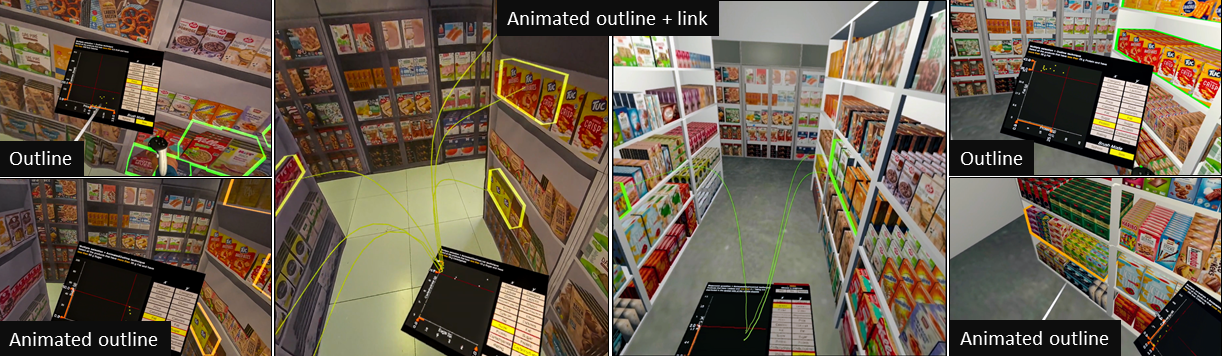}
  \caption{%
  	We leveraged highlighting techniques previously studied to perform \bl tasks in VR and AR. In our user study, users manipulated a virtual tablet to search for products of interest in a synthetic supermarket environment. Center: Visual links are displayed from the data points previously brushed by the user, connecting to the products on the shelves. Top left and right: The \textit{outline} technique highlights the product silhouette. Bottom left and right: The \textit{animated outline} technique shows a gradient hue variation (green to orange) over time. AR snapshots are shown in the left two columns, while VR snapshots are in the right two columns.%
  }
  \label{fig:teaser}
}

\graphicspath{{figs/}{figures/}{pictures/}{images/}{./}} % where to search for the images

\usepackage{tabu}                      % only used for the table example
\usepackage{booktabs}                  % only used for the table example
\usepackage{lipsum}                    % used to generate placeholder text
\usepackage{mwe}                       % used to generate placeholder figures

\usepackage{mathptmx}                  % use matching math font

\begin{document}

%%%%%%%%%%%%%%%%%%%%%%%%%%%%%%%%%%%%%%%%%%%%%%%%%%%%%%%%%%%%%%%%
%%%%%%%%%%%%%%%%%%%%%% START OF THE PAPER %%%%%%%%%%%%%%%%%%%%%%
%%%%%%%%%%%%%%%%%%%%%%%%%%%%%%%%%%%%%%%%%%%%%%%%%%%%%%%%%%%%%%%%

\firstsection{Introduction}

\maketitle

Traditional \bl techniques are used in visual data analysis to connect elements across data views using visual highlighting~\cite{Buja1991Interactive}. Such visual highlighting techniques rely on visual attributes such as color, outline, or labels. For a more precise linking between elements, there are other approaches using visual links or motion cues. 

Physical environments are naturally cluttered, and humans struggle to find and identify relevant objects.
Thus, such environments are likely to benefit from \bl. In particular, \emph{situated analytics}~\cite{Thomas2018} helps in decision-making by presenting information visualizations about the physical environment in AR. These applications need effective \bl techniques that link digital information to physical referents while preserving the perceptual characteristics of the real world~\cite{Erickson2020}.

While previous work focuses on \bl for 2D interfaces, there is only limited work on the integration of 3D environments using virtual reality (VR) or augmented reality (AR)~\cite{Ens2021}. Recently, Doerr et al.~\cite{Doerr2024VisHigh} evaluated situated visual highlighting techniques in a simulated \bl context using VR. The use of VR to simulate AR simplifies the experiment and ensures consistent image quality~\cite{Lee2013Effects, Grandi2021Design, lacoche2022arvrsimulation}.

However, a synthetic world might not match the richness and dynamics of the real world, and there may be perceptual or behavioral differences in VR compared to AR~\cite{Willet2017EmbeddedData}. Therefore, the extent to which the findings can be transferred from VR to AR remains an open question. Moreover, Doerr et al. investigated fundamental highlighting techniques, from which they extracted design guidelines. However, their paper does not cover a refinement of the highlighting based on these guidelines.

In this paper, we partially replicate, refine, and extend the study of Doerr et al.~\cite{Doerr2024VisHigh}. While the tasks of the original study were largely kept, we refined the highlighting techniques based on the existing design guidelines. Most importantly, we introduce an additional AR condition to supplement the VR condition of the original study (see~\autoref{fig:teaser}). 

In doing so, we take advantage of the fact that consumer head-mounted display (HMD) technology now offers video see-through (VST) functionality in addition to an opaque VR mode. Hence, we are able to compare the highlighting techniques of the original study using a fully synthetic model under both VR and AR conditions while using the same HMD.

With our study, we expect to obtain more robust insights into the visual perception of situated analytics problems, which \bl techniques work best for a typical situated analytics use case, as well as the fundamental feasibility of simulating AR with a VR display.

\vspace{1mm}
In summary, this paper makes the following contributions:
\begin{itemize}
\item 
A user study (N=40) to assess \bl in VR and AR using a synthetic situated analytics environment through a VST device 
\item 
Implications of AR simulation in VR on visual guidance research, particularly for situated \bl
\item 
A replication-and-extension  study that reveals interesting findings on the validity of AR simulation in VR
\end{itemize}

% ---------------------------------------------------------------------------
\section{Background and related work}
% ---------------------------------------------------------------------------

We provide some background in situated analytics (Section~\ref{sec:21}) and \bl (Section~\ref{sec:21}), the two areas that we intend to combine in our work. We also discuss the relation of our work to visual guidance research with an HMD (Section~\ref{sec:23}).

\subsection{Situated Analytics}
\label{sec:21}

Situated analytics was defined by ElSayed et al.~\cite{elsayed2015situatedanalytics} as a combination of visual analysis with the surrounding world using AR technology~\cite{Thomas2018}. Shin et al.~\cite{sungbok2024SurveySita} structure the field based on data, visualization, platform, physical location, and analytics processes. 
%They consider the surrounding data of the close objects (referents) to support analysis through visualizations that are situated on or near the referents~\cite{sungbok2024SurveySita}. 
These situated visualizations allow interactions within real-world contexts since they connect the referent with the users~\cite{Qian2024makeInteractionSituated}. 

However, enabling seamless integration of several referents and interaction modalities is a challenge for AR technologies~\cite{Sousa2023}. While the benefits of immersive virtual environments for visualizing complex spatial data can be striking, the advantages of AR can be more difficult to manifest~\cite{lacoche2022arvrsimulation}. One of the main reasons is that situated analytics is context-dependent. Several situated visualization researchers have investigated sense-making on human trajectories~\cite{luo2023pearl}, biomechanical simulations~\cite{yu2024persival}, sports training~\cite{lin2021basket}, smart objects~\cite{ye2023Proobjar} and smart objects~\cite{fleck2023ragrug}.

Even when applications rely on established design patterns for situated visualizations~\cite{Lee2024DesignPatterns}, humans do not necessarily focus on the regions of interest intended by researchers~\cite{tong2019actionUnits}. As a consequence, some work proposed region selection~\cite{shi2023Gazeselection} and effective guidance~\cite{Doerr2024VisHigh} for situated analytics contexts. 

\subsection{\Bl}
\label{sec:22}

\Bl is a standard technique in visual analysis and is related to multiple coordinated views~\cite{baldonado2000cmv}, i.e., patterns that connect data in two or more views. While brushing means selecting the data of interest, linking refers to highlighting related data in complementary views. In conventional desktop visualizations, highlighting is usually applied by altering the appearance of data points with visual encoding such as colors, glyphs, animations, or lines~\cite{Griffin2014HighlightGeovisComparison, Koytek2018MyBrush}. 

Although \bl is widely used on conventional 2D interfaces, its use in AR/VR is still an open research question~\cite{kim2025pinchcatcher}. A situated analytics survey by Shin et al.~\cite{sungbok2024SurveySita} shows that \bl techniques have rarely been evaluated in the context of a physical environment.  Previous studies on brushing focused on special aspects of real-world interaction, such as the selection of occluded objects~\cite{Sidenmark2020selectionVR, Wu2023AcquisitionOclussion} or the clutter resulting from linking of dense entities~\cite{Prouzeau2019VisualLink}. Doerr et al.~\cite{Doerr2024VisHigh} compared four highlighting techniques (color, outline, links, and arrows) in a VR supermarket shopping context. 

Efficient highlighting needs to minimize visual clutter while keeping a \emph{high contrast}~\cite{Lei2023ARVisualSalience}. However, visuals are traditionally geometric shapes that can induce side effects such as attention tunneling~\cite{syiem2021attentionTunnelingAR}. Likewise, the use of \emph{temporal cues} such as motion and flickering~\cite{Sutton2024FlickerVisualGuidance} offers an effective visual search in real-world contexts. Visual highlighting also has to be able to guide users to targets \emph{outside the field of view}~\cite{Assor2024NonVisibleSita}.

Mahmood et al.~\cite{mahmood2018multipleviewsAR} explored the coordination of multiple visualizations embedded in surfaces. Various studies applied \bl to hybrid user interfaces~\cite{Hubenschmid2021STREAM,langner2021marvis,surveyXVR2022,quijano2024BrushandLink} or to physical proxies that represent virtual content~\cite{satriadi2023Proxsituated}. However, the limited field of view of optical see-through devices makes it difficult to use these designs in AR. Video see-through displays can potentially mitigate this problem with a larger field of view. Borowski et al.~\cite{Borowski2025dashspace} followed this approach and proposed a toolkit to design visual analysis on demand in AR. Similarly, our prototype uses an HMD with video see-through features to enable exploration either in AR or in simulated AR (using VR). 

\subsection{Visual guidance}
\label{sec:23}
\label{sec:visualguidance}

Visual guidance is used to direct users to areas of interest in the environment. The effectiveness of guidance is often demonstrated in 360$^\circ$ videos~\cite{grogorick2018vgmethods, wallgrun2019vrexperience,tong2019actionUnits}. Here, information can be lost due to the lack of free navigation in videos~\cite{Gutkowski2021tourvideoAR}. Therefore, different strategies have been proposed to drive the user's attention, like displaying geometric cues and modifying the visual content. For example, Lin et al.~\cite{lin2017assistingvideos} used arrows as a hint to support immersive video navigation. Similarly, rectangles and circles have been used to locate out-of-view objects in immersive environments~\cite{Gruenefeld2018haloswedge}. 

Mixing multiple visual cues can be beneficial as well. Fox et al.~\cite{Fox2023VisualCuesLowVision} used pointers and outlines to alert people with visual impairments. Some methods modify the real content of the AR displays with respect to saturation, contrast, blurriness~\cite{sutton2022saliencyVG} or brightness~\cite{Yokomi2021VGBrightness}, but these approaches can potentially compromise the context. We only use overlaid virtual elements for visual guidance. 

% ---------------------------------------------------------------------------
\section{Study design}
% ---------------------------------------------------------------------------

The main objective of this work is to discover how accurately people perform situated \bl in a real environment compared to a simulated one. To answer this question, we extend the work of Doerr et al.~\cite{Doerr2024VisHigh}, which investigates the visual highlighting of products on shelves in a supermarket. For best comparability, we kept the original tasks of their study, but refined the highlighting techniques based on their results and recommendations. In order to compare AR and VR, we performed the study in VST-AR and VR, using an HMD that is able to support both modes. In contrast to the original, we were interested not only in the performance of the highlighting techniques but also in the differences between AR and VR of a visually identical supermarket environment.

\subsection{Scenario and dataset}
\label{sec:scenario:dataset}

A supermarket is an information-rich environment~\cite{Bowman2003} that is highly familiar to wide audiences. This property makes it a popular use case for situated analytics studies~\cite{buschel2018RealityBasedRetrieval, elSayed2024SAProcessMantra, Doerr2024VisHigh, Qian2024makeInteractionSituated}. Since user studies are difficult to perform in the public setting of a real supermarket, previous research relied on small-scale mock-ups of individual shelves~\cite{elsayed2015situatedanalytics,Sousa2023} or performed the experiment entirely in VR. We target a supermarket model that has as few visual differences as possible between VR and AR. Therefore, we adopted the approach of building the supermarket environment from scratch, using a digital-first strategy: The supermarket model was first modeled digitally and then physically built in the image of the model.

To do so, we first scraped metadata of more than 500 products from a regional supermarket website, including product name, price, description, and high-resolution images. Second, we selected 263 products sold in box-shaped containers, which fit our process of shelf placement. In comparison, Doerr et al.~\cite{Doerr2024VisHigh} used 98 products. Third, we obtained nutritional information matching the product from online databases, resulting in 10 attributes per product (\autoref{fig:teaser}). Finally, we sorted the products into 11 categories to inform the shelf layout. 
 
\begin{figure}
    \centering
    \includegraphics[height=38mm]{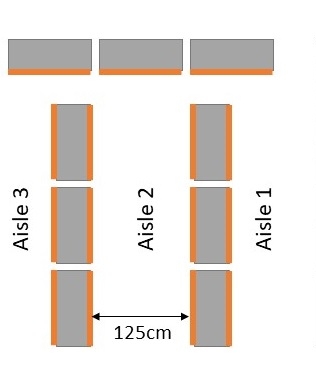}
    \includegraphics[height=38mm]{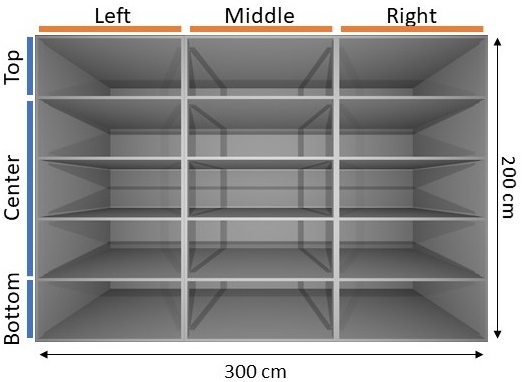}
    \caption{We built a small supermarket with three aisles, occupying approximately 25 m². Left: Structure of the supermarket model. Right: Shelf model, with the spatial arrangement for spatial judgment tasks.}
    \label{fig:aisles}
\end{figure}

Each product was modeled as a box in Blender, and the product image was applied as a texture. The dimensions of the boxes were manually adjusted to match the physical products. We organized the products into shelves using the rule that no more than three categories are allowed per shelf. The shelves are 100 cm wide and 200 cm high, with five boards each. We mimicked a typical product arrangement in a supermarket by repeating identical products side by side so that each board is filled with a maximum of five distinct products. We created 15 shelves, which were arranged in five groups consisting of three shelves next to each other, leading to a final layout consisting of three aisles (\autoref{fig:aisles}).

For the physical model, we acquired 15 steel shelves with dimensions 100$\times$200 cm from a local department store and arranged them in the chosen layout. The front face of each shelf was covered with a poster of the products printed in high resolution (20,000$\times$40,000 pixels). Using actual product boxes on the shelves would have increased the realism, but we preferred posters for multiple reasons: First, posters made the appearance of the AR supermarket more similar to the VR supermarket, because physical packaging materials and local illumination conditions did not affect the look. Second, users could not move the products, which would disrupt the registration of the real world with the digital twin. Third, highlighting techniques could exclusively focus on the front face of the box. Fourth, the effort of building (and changing) the supermarket was significantly reduced without compromising the content: Only the products' frontal appearance had to be modeled.
For a more detailed discussion of the limitations of this approach, please refer to \autoref{sec:limitations}.

\subsection{Situated visualization}

\begin{figure*}[tb]
    \centering
    \includegraphics[width=\textwidth] {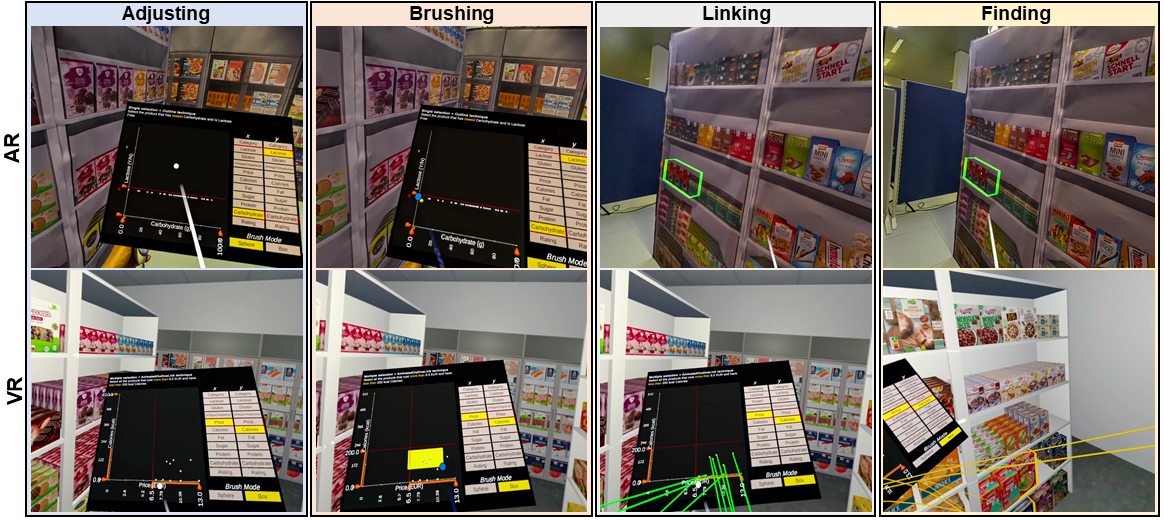}
    \caption{\label{fig:SelectionSequence}
        Brushing on a situated scatterplot in our user study. The top row shows the steps to solving a single selection task using \outline in AR, and the bottom row, a multi-selection task using \link in VR. From left to right: First, the user adjusts the data dimensions and filters to meet the task requirements. Second, they brush the data points by selecting one-to-one or delimiting regions. Third, once the data points are selected, linked referents are highlighted. Fourth, the user examines the environment to find the highlighted products, confirming by pointing a ray and pressing the trigger on the controller. Link colors are animated by transitioning from green to orange and back.}
\end{figure*}

Similar to the original study by Doerr et al.~\cite{Doerr2024VisHigh}, we used an overview visualization on a virtual tablet held in the user's hand. The overview visualization shows a 2D scatterplot and supports various touch interactions, such as filtering or brushing of the data points (\autoref{fig:teaser}, center). Each data point refers to one product. We let the user map arbitrary data attributes to the $x$- and $y$-axes of the scatterplot. Each axis is decorated with two sliders at its ends to allow the filtering of data points. Most importantly, the scatterplot lets the user brush the data points, either by tapping on individual data points sequentially or by dragging a rubber band rectangle to select the data points inside (\autoref{fig:SelectionSequence}).

\subsection{Highlighting techniques}

Visual highlighting techniques require coherent strategies to combine visual cues from the virtual and real world, enabling users to perceive areas of interest correctly. The study by Doerr et al.~\cite{Doerr2024VisHigh} investigated four basic highlighting techniques: using color overlays (thin outlines around the object and full object coloring), arrows, and visual links (leader lines). They found that the thin outlines were useful but often too inconspicuous, while the full object coloring occluded the object details. The arrows were animated, which was rated as too busy by test subjects. The visual links were found to be useful for guiding toward out-of-view objects, but otherwise were too easily overlooked. The paper concludes that outlines combined with visual links may be a potent highlighting technique.

Consequently, we designed three advanced highlighting techniques (\autoref{fig:teaser}) that pick up the lessons learned. We choose a fat outline for increased visibility, with or without additional visual links. We excluded the arrows and the full color overlay to avoid the disadvantages mentioned above.

\paragraph{\normalfont\textbf{Outline (\outline})} This technique emphasizes only the silhouette of an object, while allowing the interior surface to remain visible ~\cite{Sidenmark2020selectionVR}. Following the design recommendation of Doerr et al. and the desire for a more suitable \textit{high contrast}, we implemented the outline as a fat green line. However, including a fat contour could occlude the next objects, even more so when the object arrangements are tied to each other.

\paragraph*{\normalfont\textbf{Animated outline (\animated})} Using a single, static color might interfere with the colors of the background, like a green outline on a green product box.
Motivated by the feasibility of \textit{temporal visual cues} for visual guidance~\cite{Sutton2024FlickerVisualGuidance}, we therefore opted for an animated outline that avoids this problem by smoothly changing the highlighting color between green and orange over time.
The animation runs continuously and is synchronized for all highlighted products to minimize temporal clutter.

\paragraph*{\normalfont\textbf{Animated outline + linking (\link})} While an animated outline ensures high contrast with the background, it does not guide to off-view referents.
Hence, we added visual links from the data points in the scatterplot to the corresponding referents~\cite{Prouzeau2019VisualLink}. The links are drawn with a curved line and smoothly update for tablet motions (and, with it, the data points). For \textit{out-of-view} products, the line has a visible display portion on the sides to indicate the connection to these referents. Compared to Doerr et al.~\cite{Doerr2024VisHigh}, we increased the curvature of the links and smoothed the transitions caused by the movement of the scatterplot from the activity of natural users (\autoref{fig:teaser}).

\subsection{Tasks}
\label{sec:tasks}

To ensure comparability, we adapted the tasks of Doerr et al.~\cite{Doerr2024VisHigh}, which involve appropriate interactions for situated analytics~\cite{Qian2024makeInteractionSituated}.

\textbf{Single Selection:} 
This task requires locating one single referent, e.g., \textit{``Select the product that has the highest fat and has gluten.''}

\textbf{Multiple Selections:}
This task requires locating multiple referents, e.g., \textit{``Select all the products that have more than 36 g fat and have less than 50 g sugar.''}

\textbf{Spatial Judgment:}
This task requires the user to judge the spatial arrangement of groups of referents. The user must indicate whether the highlighted referents are located on the left, center, right, and the top, middle, or bottom of the shelves (e.g., \textit{``Products that have less than 3.3 g fat and free lactose are distributed on the top side of the left shelves.''}) The spatial arrangement is taken from the perspective of the participant~(\autoref{fig:aisles}, right); no mental rotation is required. This type of task was chosen to explore whether participants could handle challenging real-world situations.

Each multi-selection task requires 10 products, and the chosen referents were distributed uniformly across the shelves.

% ---------------------------------------------------------------------------
\section{Experiment}
% ---------------------------------------------------------------------------

We conducted a within-subjects study with \textit{display context} (VR or AR) and \textit{highlighting technique} (\outline, \animated, or \link) as independent variables, resulting in $2\times3=6$ conditions. The experiment was carried out in a 5$\times$3 m wide area in the middle aisle. The participant was asked to stand in the center of the aisle. We included referents from the three shelves to the left and the three shelves to the right. These six shelves contained 101 referents. The remaining seven shelves were used only as decoration and were not included in our study. 

From the user's position, only one of the opposing sides of the aisle was visible at any time. The user had to turn the head or body to face the other side. Walking was not necessary to solve the tasks.

The VR and AR conditions were performed at the same location, and the supermarket was shown on the same scale. However, in the VR condition, the user's entire field of view of was filled with the virtual supermarket, while the video see-through was turned off. Thus, the VR and AR conditions only differed in the presentation of the supermarket model.

\subsection[]{\added{Hypotheses}}
\label{sec:hypotheses}

We had the following three main \added{hypotheses} for our experiment.

\noindent\textit{\added{H1}: Similar results will be found under VR and AR conditions.}
Previous studies argue that VR experiments can be representative of AR experiments. However, there is no systematic evidence on how far this transfer of findings from VR to AR goes. We purposely made the VR and AR conditions as similar as possible to uncover any remaining differences.

\noindent\textit{\added{H2}: \animated will perform better than \outline.}
Several works benefit from the use of animation to compare trends, provide visual guidance~\cite{ChromaGazer2025Tosa}, and direct attention in virtual environments. We expected that completion times and error rates would benefit from animated colors.

\noindent\textit{\added{H3}: \link will give the best results overall.} 
We expected that the additional guidance provided by the visual links would facilitate \bl when out-of-view referents are involved.

\subsection{Participants}

We recruited 42 graduate students from our university campus, but \added{after observing the recording data,} two participants were removed from our analysis due to inattention \added{caused by excessive delay between trial initiation and task execution} and a technical problem. The final set of 40 participants (17 female and 23 male) ranged in age from 18 to 44 (eight participants from 18 to 24, 30 participants from 25 to 34, and two participants from 35 to 44). All had normal or corrected-to-normal vision and, if necessary, wore glasses in combination with the HMD. Their reported previous experience with an HMD was no/low for 15 participants and high for ten participants. Of the 40 participants, five reported having no experience with 3D computer games, and three were left-handed.

\subsection{Apparatus and implementation}

We used a Meta Quest 3 (running Horizon OS 72.0) with touch controllers and Unity 3D (version 2022.3.34f1) with the Meta XR Interaction SDK (68.0.1) to implement the application. The device has a horizontal field of view of approximately 110 degrees. For the study, we ran the application standalone and set the display to have at least 72 fps. In addition, the environment calibration is performed using fixed virtual anchors on the floor squares.
For the AR condition, we used the pass-through capability of the device. In addition, a virtual tablet of 40 cm by 31 cm was presented, attached to the controller held in the non-dominant hand. The tablet showed instructions, tasks, and a scatterplot visualization. Brushing on the tablet was implemented by ray-casting from the controller held in the dominant hand. We assigned the trigger and grip buttons on the controller to highlight and de-highlight, respectively. 

\subsection{Measures}
\label{sec:measures}

We defined the following primary measures that apply to all 18 tasks featured in our experiment: 

\textit{(1) Completion time.} We measured the time interval from task start to the selection of an answer. 

\textit{(2) Linking time.} We measured the time that the participant's gaze was fixed on the environment, instead of the scatterplot. 

\textit{(3) Error rate.} We determined binary accuracy: In the selection tasks, we determined whether or not the participant was entirely correct in finding the referents on the first attempts. In spatial judgment tasks, the answer is binary. The error rate was directly computed as the fraction of correct answers.

Additionally, we measured \textit{workload}, using NASA-TLX~\cite{Hart2006NasaTlx}, \textit{usability}, measured using Umux-lite~\cite{lewis2013umux}, and \textit{presence}, measured using the Igroup Presence Questionnaire (IPQ)~\cite{IPQ2024}. All measures are averaged over the number of repetitions.

\subsection{Procedure}

\added{Our user study protocol was approved by the Committee for Responsibility in Research (Ethics Committee) at the University of Stuttgart.} A Latin square design counterbalanced the order of the conditions. Each participant performed six sessions, where they started using one of the three highlighted techniques in the VR condition and later continued in a similar order in the AR condition (or vice versa). Within a condition, the task order was fixed, consisting of one single, multi-, and spatial judgment task per technique. We created six tasks per task type (18 in total) to prevent learning effects.
Before the actual experiment, we performed a pilot study (N=5). The results allowed us to improve the clarity of the instructions given to the participants and to accelerate the animation sequences.

The experiment lasted approximately one hour. In total, the experiment collected data from 40 participants $\times$ 2 viewing contexts $\times$ 3 techniques $\times$ 3 task types = 720 trials.
After an introduction and signing of the consent form, participants were introduced to the virtual tablet and interaction options. In addition, we showed the supermarket and explained the spatial arrangement inside a shelf (top, center, bottom) and in a group of three shelves (left, middle, and right), used in the spatial judgment tasks.
Next, the experimenter calibrated the environment and adjusted the controller interface to match the handiness of the participant so that they could hold the virtual tablet with the non-dominant hand and interact with the dominant hand. The participants then put on the headset and grabbed both controllers.
For each condition, the environment was recalibrated, and the participants went through a training phase, followed by the main trials, including questionnaires and a short semi-structured interview. Finally, participants were compensated with 14 EUR.

During \emph{training} (5-10 minutes), the experimenter explained the tablet interface without active tasks.
The participants could familiarize themselves with the features, try the selection of data points using both brushing and filters, and observe the highlighting technique applied to the chosen referents. This training phase used a modified dataset. 

In the \emph{main trials} (30-40 minutes), the order of conditions was counterbalanced, but the order of tasks in one condition was fixed: single, multi-selection, and spatial judgment tasks. Before each trial, a text informing about the task was shown on the tablet. Once the participants pressed the start button, the scatterplot was displayed. For the selection tasks, the participants had to select all the highlighted referents by pointing a ray towards the referent and confirming with a trigger button. In the case of the spatial judgment task, the tablet showed three options, ``yes'', ``no'' and ``maybe''. We intentionally included the option ``maybe'' to reduce guessing, in accordance with Doerr et al.~\cite{Doerr2024VisHigh}. Once a task was completed, all highlighting was reset, and the participants were required to return to the center of the aisle. After the three highlighting tasks (\autoref{sec:tasks}), the participants were invited to complete the workload questionnaire. At the end of the condition, they completed the usability and presence questionnaires. After the VR or AR parts of the experiment, they rated the three highlighting techniques. 

Finally, participants filled out a demographic questionnaire, and we conducted short \emph{semi-structured interviews} (\textless5 minutes). Here, we asked some questions about the overall experiment, including preferred techniques and subjective task performance.

\begin{figure}[tb]
    \centering
    \subfloat {\includegraphics[height=1.5cm] {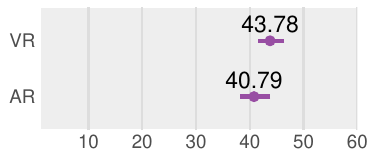}}
    \quad
    \subfloat {\includegraphics[height=1.5cm] {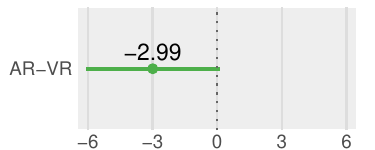}}
    \\
    \subfloat {\includegraphics[height=1.5cm] {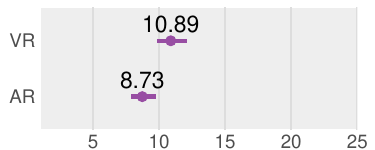}}
    \quad
    \subfloat {\includegraphics[height=1.5cm] {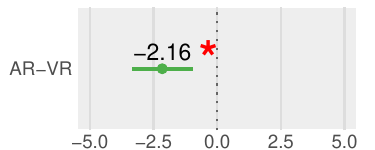}}
    \caption{\label{fig:Overall_times_across_conditions}
           (left) \textcolor{timingColor}{Mean Completion Time} (top) and \textcolor{timingColor}{Mean Linking Time} (bottom) in seconds for all conditions and tasks. (right) \textcolor{pairwise}{Pairwise differences}. Error bars represent 95\% bootstrap confidence intervals. Evidence of differences is marked with an asterisk \textcolor{red}{\textbf{*}}. The further away from zero and the tighter the CI, the stronger the evidence is.}
\end{figure}

\begin{figure}[tb]
    \centering
    \subfloat {\includegraphics[height=1.5cm] {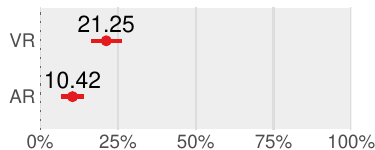}}
    \quad
    \subfloat {\includegraphics[height=1.5cm] {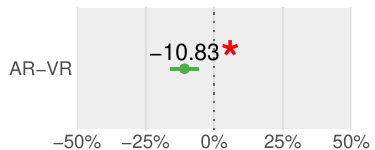}}
    \\
    \subfloat {\includegraphics[height=1.5cm] {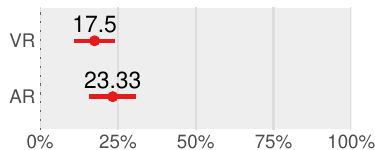}}
    \quad
    \subfloat {\includegraphics[height=1.5cm] {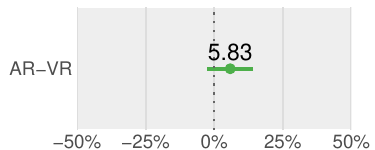}}
    \caption{\label{fig:Overall_errorRate_conditions}
    (left) \textcolor{errorRate}{Mean error rate} in \% for selection tasks (top) and for spatial judgment tasks (bottom), for all conditions and tasks. (right) \textcolor{pairwise}{Pairwise differences}. Error bars representing 95\% bootstrap confidence intervals. Evidence of differences is marked with an asterisk \textcolor{red}{\textbf{*}}. The further away from 0\% and the tighter the CI, the stronger the evidence is.}
\end{figure}

% ---------------------------------------------------------------------------
\section{Results}
% ---------------------------------------------------------------------------

Following recommendations for establishing a modern evaluation methodology~\cite{theNewStatistics2014, dragicevic2016fair}, we report our inferential statistics using \textit{interval estimation} instead of p-values. \textit{Confidence intervals} (CI) were calculated using \textit{bias-corrected and accelerated bootstrapping}~\cite{efron1987better}. CI allow for more meaningful interpretations. \added{Pairwise \ciicon } CI  that do not overlap with zero correspond to p-values smaller than the significance level, providing the same evidence of  differences as in traditional null-hypothesis significance testing. \added{On the other hand, crossing with zero indicates no statistically significant difference.} Hence, no p-values are reported, but they can be obtained from CI results~\cite{Krzywinski2013pvalues}.
\added{Source code, } data collected, and analysis scripts are available as supplementary material\footnote{\url{https://github.com/cquijanoch/SituatedBrushingAndLinking}}.

\subsection{Overall results across conditions}
\label{sec:overall:condition}

\autoref{fig:Overall_times_across_conditions}~(top) shows \textit{completion time} for all tasks. Participants took less than one minute to complete each task, being faster in AR (40.79 s) than in VR (43.78 s). No evidence was found that completion times differ significantly between display conditions.

\autoref{fig:Overall_times_across_conditions}~(bottom) shows mean \textit{linking time} for all tasks. Participants spent less than 15 seconds observing outside the tablet for each task. Mean times are shorter for AR (8.73 s) than for VR (10.89 s), indicating that AR is 2.16 s faster than VR.

\autoref{fig:Overall_errorRate_conditions}~(top) shows the mean \textit{error rate} for selection tasks. There is clear evidence that AR is less error-prone (10.42\%) than VR (21.25\%), with an estimated difference of 10.83\%.

\autoref{fig:Overall_errorRate_conditions}~(bottom) \added{shows the mean \textit{error rate} for spatial judgement tasks.}  While AR is more error-prone (23.33\%) than VR (17.5\%) on average, there is no evidence that mean error rates differ between conditions for spatial judgment tasks.

Overall, the findings do not align with \added{\textbf{H1}}: While completion time and spatial judgment error rate do not show meaningful differences between conditions, AR reduces linking time and is less error-prone than VR for selection tasks. 

\begin{figure}[tb]
    \centering
    \subfloat {\includegraphics[height=3cm] {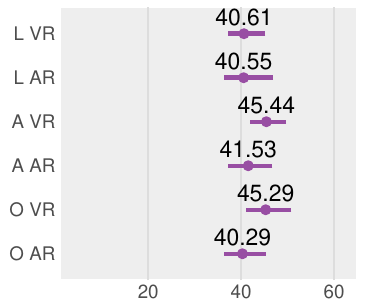}}
    \quad
    \subfloat {\includegraphics[height=3cm] {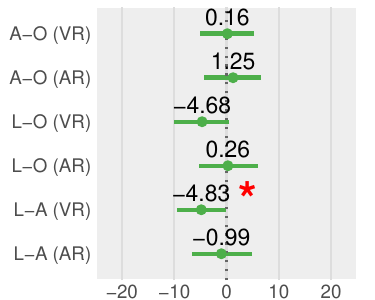}}
    \\
    \subfloat {\includegraphics[height=3cm] {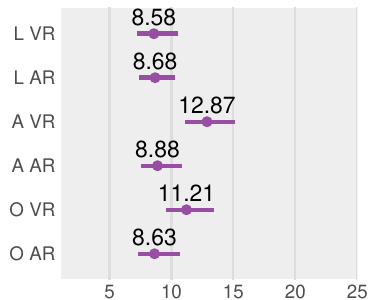}}
    \quad
    \subfloat {\includegraphics[height=3cm] {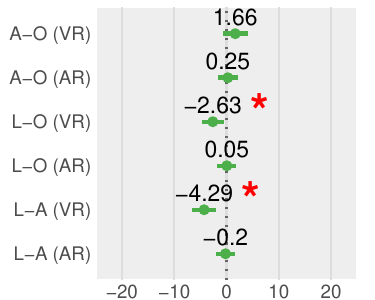}}
    \caption{\label{fig:Time_perTechnique}
           (left) \textcolor{timingColor}{Mean completion time} (top) and \textcolor{timingColor}{Mean linking time} (bottom) in seconds per technique. (right) \textcolor{pairwise}{Pairwise differences}. Error bars represent 95\% bootstrap confidence intervals. Evidence of differences is marked with an asterisk \textcolor{red}{\textbf{*}}. The further away from zero and the tighter the CI, the stronger the evidence is.}
\end{figure}

\begin{figure}[htb]
    \centering
    %\subfloat {\includegraphics[height=3cm] {figures/Results/plot_AccuracyTechnique_S_CI.pdf}}
    %\quad
    %\subfloat {\includegraphics[height=3cm] {figures/Results/plot_AccuracyTechniqueDiff_CI_S.pdf}}
    %\\
    \subfloat {\includegraphics[height=3cm] {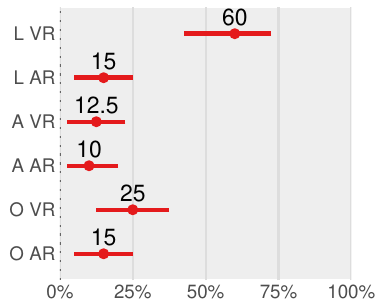}}
    \quad
    \subfloat {\includegraphics[height=3cm] {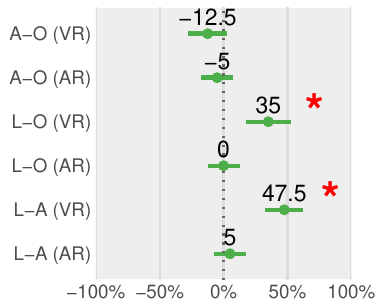}}
    \\
    \subfloat {\includegraphics[height=3cm] {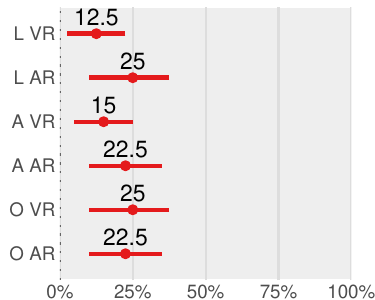}}
    \quad
    \subfloat {\includegraphics[height=3cm] {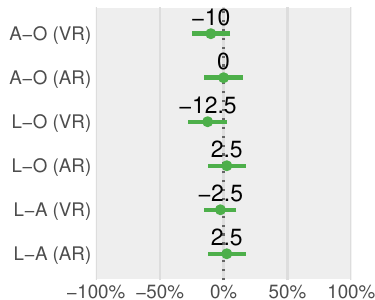}}
    \caption{\label{fig:ErrorRate_perTechnique}
           \textcolor{errorRate}{Mean error rate} in \%  for multi-selection tasks (top) and for spatial judgment tasks (bottom) per technique. (right) \textcolor{pairwise}{Pairwise differences}. All error bars represent 95\% bootstrap CI. Evidence of differences is marked with an asterisk \textcolor{red}{\textbf{*}}. The further away from 0\% and the tighter the CI, the stronger the evidence is.}
\end{figure}

\subsection{Results per technique}
\label{results:technique}

\autoref{fig:Time_perTechnique}~(top) suggests that the mean \textit{completion time} in AR is shorter for {\outline} (40.29 s), followed by {\link} (40.55 s) and {\animated} (41.53 s). However, there is no strong evidence of differences between the three techniques. For VR, \link performed the fastest (40.61 s), followed by \outline (45.29 s) and \animated (45.44 s). We found evidence of a meaningful difference between \link and \animated, with \link being faster than \animated by 4.83 s. While \link is the fastest in VR, there are no remarkable differences in AR, on average.

\autoref{fig:Time_perTechnique}~(bottom) shows that mean \textit{linking time} in AR is shortest for \outline (8.63 s), followed by \link (8.68 s) and \animated (8.88 s). However, there is no strong evidence of meaningful differences between the three techniques. For VR, \link has the fastest linking time (8.58 s), followed by \outline (11.21 s) and \animated (12.87 s). Evidence shows that \link is faster than \outline by 2.63 s and  \animated by 4.29 s. In addition, differences suggest that \outline is faster than \animated in VR. There are no considerable differences among techniques in AR.

\autoref{fig:ErrorRate_perTechnique}~(top) shows that mean \textit{error rates} in AR are lowest for \animated (10\%), followed by \link (15\%) and \outline (15\%). There is no strong evidence of meaningful differences among the three techniques. For VR, \animated is also less error-prone (12.5\%), followed by \outline (25\%) and \link (60\%). There is strong evidence that \link is less accurate than \outline by 35 percent points and \animated by 47.5 percent points. Besides, pairwise differences show that \animated is more accurate than \outline. Regarding spatial judgment accuracy (\autoref{fig:ErrorRate_perTechnique}, bottom), while there are no large differences between the techniques, on average, \outline performs better in VR.

Thus, the results align only partially with our \added{hypothesis \textbf{H2}}.
\animated is less prone to error than \outline. However, the participants spent slightly more time with \animated. For \added{\textbf{H3}}, we found that the completion time and linking time of \link are the fastest only for VR. However, \link is more error-prone for VR. Hence, \added{\textbf{H3}} is partially supported for VR.

\begin{figure*}
\centering
\subfloat[\label{fig:nasatlx:1}Mental Demand] 
{
    \includegraphics[height=3cm]{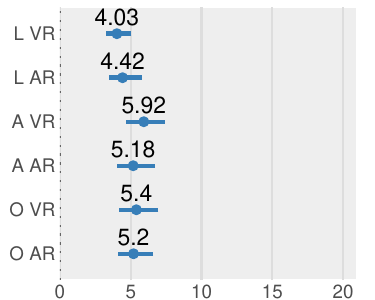}
    \includegraphics[height=3cm]{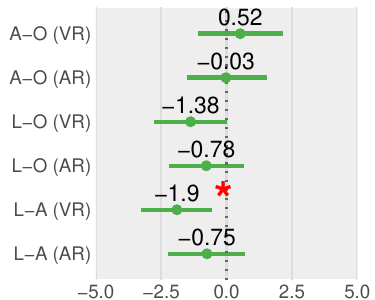}
}
\quad
\subfloat[\label{fig:nasatlx:2}Physical Demand] 
{
    \includegraphics[height=3cm]{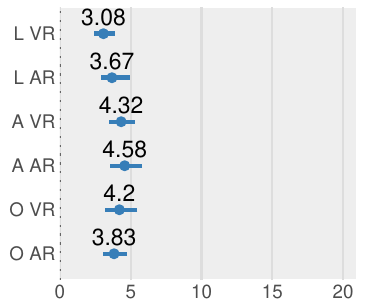}
    \includegraphics[height=3cm]{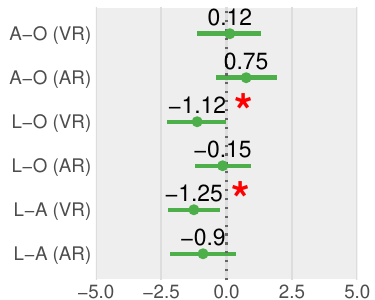}
}
\quad
\subfloat[\label{fig:nasatlx:5}Effort] 
{
    \includegraphics[height=3cm]{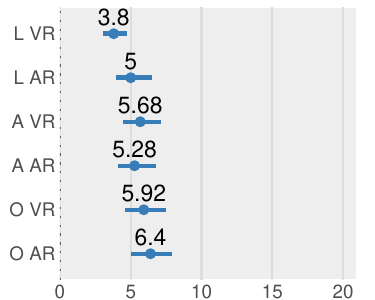}
    \includegraphics[height=3cm]{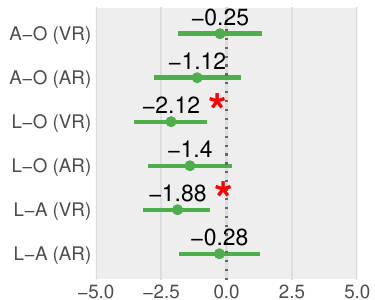}
}
\quad
\subfloat[\label{fig:nasatlx:6}Frustration] 
{
    \includegraphics[height=3cm]{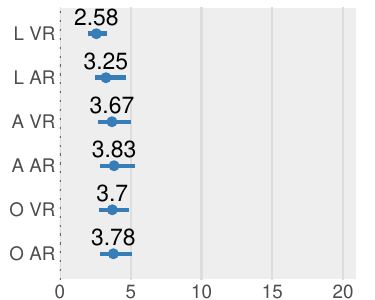}
    \includegraphics[height=3cm]{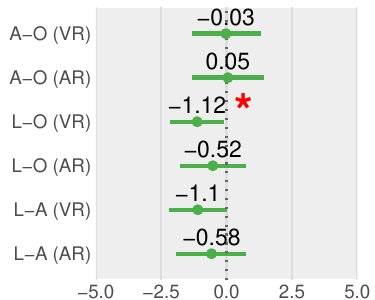}
}
\caption{Results for NASA TLX factors for all conditions and tasks (lower is better). \textcolor{workload}{Mean workload factor scores}. \textcolor{pairwise}{Pairwise differences}. Error bars represent 95\% bootstrap confidence intervals. Evidence of differences is marked with an asterisk \textcolor{red}{\textbf{*}}. The further away from zero and the tighter the CI, the stronger the evidence is. We did not find evidence of differences for temporal demand and performance factors.}
\label{fig:NasaTLX}
\end{figure*}

\subsection{Workload}
\label{sec:workload}

In addition to time and accuracy, we were interested in investigating possible differences in perceived workload between highlighting techniques using NASA-TLX. \autoref{fig:NasaTLX}  suggests evidence of differences in VR only. For mental demand, physical demand, and effort, results indicate that \link performs better than \animated. The results in~\autoref{fig:NasaTLX} also suggest that \link reduces the physical demand, effort, and frustration compared to \outline.

\begin{figure}[tb]
    \centering
    \subfloat {\includegraphics[height=1.5cm] {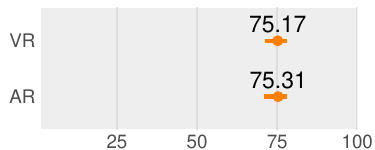}}
    \quad
    \subfloat {\includegraphics[height=1.5cm] {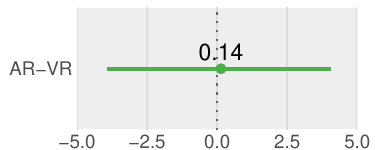}}
    \caption{\label{fig:usability}
           (left) \textcolor{usability}{Mean SUS score} evaluated by condition. (right) \textcolor{pairwise}{Pairwise differences}. Error bars represent 95\% bootstrap CI.}
\end{figure}

\subsection{Usability and presence}
\label{sec:usability:presence}

We calculated the System Usability Scale (SUS) from the Umux-Lite~\cite{lewis2013umux} filled out after the condition trials. However, as shown in \autoref{fig:usability}, the score for AR (75.31) is only slightly higher compared to VR (75.17), with no evidence of a meaningful difference.

In addition, we explored the presence factors of both conditions and found a similar ``sense of being there'' mean score of 4.05 for AR and 4.6 for VR. Exploring the presence factors, \autoref{fig:presence}~(left) shows that the sense of being physically present (SP) for VR is higher (4.26) than AR (3.92). For the subjective experience of realism (REAL), results suggest that AR is slightly higher (3.48) than VR (3.24). The involvement (INV) for VR is higher (3.81) than AR (2.11). \autoref{fig:presence}~(center) shows that there is evidence that involvement for VR is greater than AR by 1.7.

\subsection{User preferences}
\label{sec:preferences}

\begin{figure}[tb]
    \centering
    \subfloat {\includegraphics[height=2.4cm] {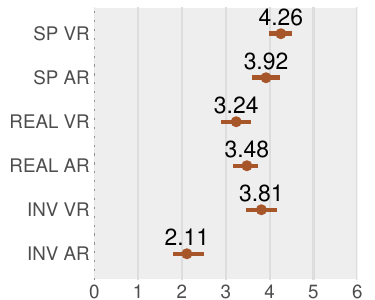}}
    \subfloat {\includegraphics[height=2.4cm] {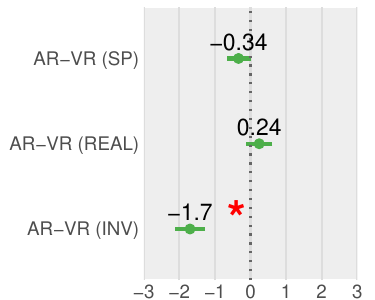}}
    \subfloat {\includegraphics[height=2.4cm] {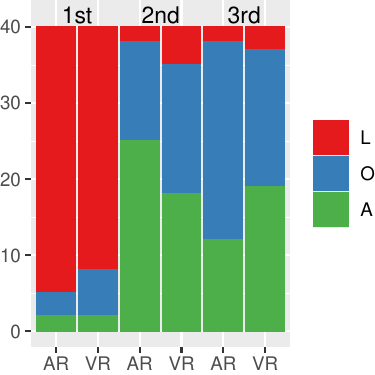}}
    
    \caption{\label{fig:presence}
           (left) \textcolor{presence}{Mean presence factor scores} (INV: Involvement, REAL: Realism, and SP: Spatial presence) for both conditions. (Center) \textcolor{pairwise}{Pairwise differences}. Error bars representing 95\% bootstrap confidence intervals. Evidence of differences is marked with an asterisk \textcolor{red}{\textbf{*}}. The further away from 0\% and the tighter the CI, the stronger the evidence is. (right) Distribution of rating choice across the highlighting techniques.}
\end{figure}

Based on our results, all participants quickly learned how to interact with the application to solve the tasks under both conditions. \autoref{fig:presence}~(right) shows participants' responses for their preferred highlighting technique by condition. Most of the participants rated \link as their first choice in AR (87.50\%) and VR (80\%).
The second choice rating shows that for AR, \animated is the most rated (62.50\%), followed by \outline (32.5\%), while for VR, \animated and \outline were rated similarly by the participants (45\% and 42.5\%, respectively).

The responses in the interview section confirmed the previous analysis: Most participants agreed that visual links (L) helped track the referent's location, avoiding searching in other places. However, we found that \link appears obtrusive. P9 commented that \textit{``the links appear extremely fast and may cause sickness''}; P12 made a similar comment. In addition, P24 added that \textit{``the popup of the links is worse in AR than in VR, in VR it is more smoothly''}. Moreover, P18 commented that \textit{``Link works good when you are selecting few data points''}. P13 emphasized that \textit{``the links distract from checking the surroundings to find the products''}. We also observed that the users looked back once the data points were highlighted and the links were displayed. This behavior happened when the participants applied the rectangle brushing mode.  

Regarding the effectiveness of color animation (A), we found that the majority of the participants did not realize the color was animated. However, there is no simple solution. Using a faster animation was suggested by three participants. For example, P17 commented that \textit{``the animation was at first not perceptible''}, and P20 that \textit{``I need to wait for the change of color''}. However, P1 mentioned that \textit{``Animation is too fast. Blinking could cause illness''}. Eight participants suggested different colors: P2, P4, P31, and P37 proposed that the animation should consider the current product and shelf colors. P27 and P33 suggested using colors such as red, blue, pink, and purple, while P11 suggested using colors by product category. In addition, we detected that the participants had difficulty finding green and orange box products, mainly in VR. P3 pointed out that \textit{``it was harder to recognize the green box because the green highlighting color in VR, maybe for this case it makes more sense the AR''}. P20 commented to like \textit{``outline better than the animated one because the color [green] contrasts with crackers [orange] and other products''}. Some participants interpreted the orange color as response feedback. P19 remarked that \textit{``after some seconds, I felt the animated colors looked like a false response''}. P34 pointed out that \textit{``color could change with the gaze, by product independently, instead of as a whole.''}.

Participants enjoyed the VR condition due to its high immersion, but also remarked that AR is very promising. P2 pointed out that \textit{``there are benefits of highlighting techniques in AR due to the display of the device''}. P15 also added that \textit{``in AR it is easier to perform the tasks, but VR looks like more fun''}. Furthermore, P7, P17, P18, and P34 complained about the HMD resolution.

% ---------------------------------------------------------------------------
\section{Discussion}
% ---------------------------------------------------------------------------

This study compared three highlighting techniques---\outline, \animated, \link---to perform situated \bl tasks in VR and AR.
Since we replicated the work of Doerr et al.~\cite{Doerr2024VisHigh} only on the conceptual level of feature and design choices, the results cannot be directly compared. Therefore, we discuss our main findings (in the light of theirs, when possible), implications, and limitations of our study. 

\subsection{Main Findings}
\paragraph*{Visual highlighting for situated \bl differs between AR and VR.} It seems that the visual recognition of the highlighting differed between AR and VR (\autoref{sec:overall:condition}). It was more clearly perceived in AR because of the noticeable visual differences between the video-see-through content and the overlays. This advantage resulted in faster and more accurate performance in AR. As long as VR cannot deliver a visual scene representation that is indistinguishable from AR, we must expect performance differences between VR and AR.

\paragraph*{\link performs faster overall but is less accurate in VR.} We assume the visual links assist in achieving the target faster in both conditions; however, the silhouette is difficult to perceive in VR. Hence, the end routing of the link mainly influences accuracy in VR compared to AR.  
The 3D rendering makes the user perceive one congruent scene in which visual cues affect all content uniformly. The silhouette of the objects was confused with the body of the objects and affected target recognition. This fact is also reflected in the workload results.  

\paragraph*{\animated is more accurate but slower than unique color.} Although the results do not show strong evidence, the completion time of \outline is lower than that of \animated in both conditions (\autoref{fig:Time_perTechnique}). However, the error rate of \animated is lower than that of \outline for finding referents (\autoref{fig:ErrorRate_perTechnique} top). We assume that this might be due to the chosen color sequence and animation speed. We deduce that, among the outline techniques, \animated is unobtrusive for visual guidance, but time-consuming.

\paragraph*{\link is most preferred by the participants.}  Similarly to Doerr et al.~\cite{Doerr2024VisHigh}, visual links are subjectively preferred. However, this technique copies the visual properties of \animated and extends it by a visual line to guide to targets, which limits its comparability. Visual link design should be investigated in more detail to make it more effective.

\paragraph*{Highlighting techniques lead to the same spatial judgment.} The spatial abilities (not discussed in detail by Doerr et al.~\cite{Doerr2024VisHigh}) that users exhibited in our experiment did not differ between AR and VR. The results suggest that there are no differences with respect to spatial judgment by display or technique condition yet. The reason could be that all techniques offer a similar level of spatial awareness. In that sense, the AR environment can be replaced by a simulated VR environment without disadvantages in the spatial abilities, as expected from the results obtained in spatial presence research. % (\autoref{sec:usability:presence}). 

\subsection{Implications and future work}

\paragraph*{Visual links cause discomfort and occlusion.} As reported in~\autoref{sec:preferences}, multiple artificial marks displayed in a frame cause occlusion and visual discomfort. For example, when brushing multiple data points at once, links add clutter to the tablet interface. Some visual links can be occluded if targets are behind the tablet or the links penetrate the image plane. We propose that visual links should be displayed with a faded-out appearance near their origin to minimize the clutter. In addition, the links could be routed away from the user's gaze direction. Another strategy is to change the origin of the visual links. We always start the visual line from the data points. Other layouts are conceivable, such as starting visual links from the closest edge of the tablet. We also speculate that visual links floating in physical space could be distracting. Especially if \link connects to an item in a different aisle, it may be better to route visual links on the floor and ceiling. Such a use of `above' and `below' has been previously suggested by Satkowski et al.~\cite{satowski2018abovebelow}. Besides, the end routing of the link matters for cluttered targets. Adjusting the ending line, which reaches the referents, to be perpendicular and thicker could ensure greater accuracy.

\paragraph*{Animated color is unobtrusive but time-consuming.} We observed that the animated highlighting color (green to orange) did not optimally consider colorimetry factors. Future studies could evaluate the effectiveness of \animated in situated \bl. For example, a recent study~\cite{ChromaGazer2025Tosa} assessed non-obtrusive animation colors for visual guidance on standard displays. However, they did not explore their performance using current VR or AR displays. In addition, following the suggestion of the participants (\autoref{sec:preferences}), the effectiveness of dynamically changing the color considering the object's own color is an opportunity for future implementations.  

\paragraph*{Outline techniques require guidance strategies.} Both outline techniques cannot highlight out-of-view targets, which limits their application in complex layouts, where targets located in other aisles are occluded (\autoref{fig:aisles}). Our experiment only considered a single aisle. In the future, a multi-aisle visualization could use a ``magic lens'' to reveal referents behind the occluders. This pattern is commonly used in situated analytics~\cite{Lee2024DesignPatterns}, but its effectiveness for \bl is currently unexplored.

\paragraph*{\Bl physical objects of interest.} The current brushing interactions focused on one direction, from the data points to the physical referents. However, in real scenarios, users can interact with the physical environment. As AR leverages the physical environment, users can acquire information about the physical objects on demand. Querying information based on visible objects in unexpected scenarios is challenging to simulate in VR, due to the isolation from the real world, e.g., retrieving information from picking objects. Further work could assess object selection of physical referents in AR~\cite{shi2023Gazeselection, quijano2024BrushandLink} to evaluate all directions of situated \bl at once.

\paragraph*{Situated \bl in VR} We noticed that subtle differences in visual realism between VR and AR can lead to performance differences in an otherwise identical task. In our experiment, AR had the advantage that highlights were more easily identified in comparison to the video background, compared to the synthetic background in VR. This observation suggests that VR simulation has its limits. Users in VR could be influenced by illusion rather than realism~\cite{Jung2021RealismPresenceVR}.

\subsection{Limitations}
\label{sec:limitations}
For the AR condition, we built physical shelves but used high-resolution printed posters for the products. As pointed out in \autoref{sec:scenario:dataset}, this procedure has several advantages in terms of costs, flexibility, and reproducibility. However, it also reduces the realism in the 3D perception of the products\added{, possibly affecting ecological validity}. We tried to limit this effect by basing the highlighting techniques on the digital 3D models\added{, keeping the spatial arrangement the same in all conditions, and not involving the need for touching products or walking,} even in the AR condition, ensuring that the techniques worked the same in VR and AR. However, four of our 40 participants mentioned the use of posters as a limitation, and we cannot completely rule out a negative effect on performance in the AR condition. 

Another limitation \added{is the use of only one HMD model,}  the Meta Quest 3, in our study. Future devices may differ with respect to the field of view, resolution, or distortions. Therefore, special care must be taken when generalizing our findings to arbitrary headsets. 

Moreover, the performance results on highlighting (linking) reported in this paper may be affected by errors or misinterpretation in the brushing phase. To avoid interrupting users after each step required in a given task, we considered \bl as a single process judged as a whole. Users who erred during brushing would have likely performed poorly on the linking part, but our experimental procedure did not explicitly split these causes. In addition, additional perceptual and cognitive metrics may be necessary to get a more complete picture of how humans operate in VR and AR.

% ---------------------------------------------------------------------------
\section{Conclusion}
% ---------------------------------------------------------------------------

This work evaluates the performance of highlighting techniques for situated analytics. We revisit insights from a previous study by Doerr et al.~\cite{Doerr2024VisHigh} and address the topics left open in that work. Specifically, we focus on refined highlighting techniques that satisfy the design guidelines put forward by Doerr et al., and we investigate how perception of the \bl changes in a VST-AR setting taking place in a physical supermarket as opposed to a mere virtual one. Our results suggest that VST-AR performs better in task performance and accuracy overall\added{. Likewise, the animated linking performs faster but less accurately in VR only, whereas the use of animated colors improves accuracy compared to static color, albeit at the expense of time.} 

%% if specified like this the section will be omitted in review mode
\acknowledgments{%
	We wish to thank the anonymous reviewers for their comments. Thanks are also due to Fairouz Grioui and Sanan Akther for their participation in setting up the supermarket.
    This work was supported by the Alexander von Humboldt Foundation funded by the German Federal Ministry of Research, Technology and Space, German Research Foundation DFG (495135767), and Austrian Research Funds FWF (I5912).
}

\bibliographystyle{abbrv-doi-hyperref}

\bibliography{template}

@STRING{CHI = {ACM Conference on Human Factors in Computing Systems}}

@string{ETRA = {Symposium on Eye Tracking Research and Applications (ETRA)} }

@article{Doerr2024VisHigh,
author = {Doerr, Nina and Lee, Benjamin and Baricova, Katarina and Schmalstieg, Dieter and Sedlmair, Michael},
title = {Visual Highlighting for Situated Brushing and Linking},
journal = {Computer Graphics Forum},
volume = {43},
number = {3},
pages = {e15105},
doi = {https://doi.org/10.1111/cgf.15105},
year = {2024}
}

@article{efron1987better,
  title={Better bootstrap confidence intervals},
  author={Efron, Bradley},
  journal={Journal of the American Statistical Association},
  volume={82},
  number={397},
  pages={171--185},
  year={1987},
  publisher={Taylor \& Francis},
  doi={10.2307/2289144}
}

@ARTICLE{Griffin2014HighlightGeovisComparison,
  author={Griffin, Amy L. and Robinson, Anthony C.},
  journal={IEEE Transactions on Visualization and Computer Graphics}, 
  title={Comparing Color and Leader Line Highlighting Strategies in Coordinated View Geovisualizations}, 
  year={2015},
  volume={21},
  number={3},
  pages={339-349},
  doi={10.1109/TVCG.2014.2371858}}

@inproceedings{Qian2024makeInteractionSituated,
author = {Zhu, Qian and Wang, Zhuo and Zeng, Wei and Tong, Wai and Lin, Weiyue and Ma, Xiaojuan},
title = {Make Interaction Situated: Designing User Acceptable Interaction for Situated Visualization in Public Environments},
year = {2024},
isbn = {9798400703300},
publisher = {Association for Computing Machinery},
address = {New York, NY, USA},
url = {https://doi.org/10.1145/3613904.3642049},
doi = {10.1145/3613904.3642049},
booktitle = {Proceedings of the 2024 CHI Conference on Human Factors in Computing Systems},
articleno = {196},
numpages = {21},
location = {Honolulu, HI, USA},
series = {CHI '24}
}

@inproceedings{buschel2018RealityBasedRetrieval,
author = {B\"{u}schel, Wolfgang and Mitschick, Annett and Dachselt, Raimund},
title = {Here and Now: Reality-Based Information Retrieval},
year = {2018},
isbn = {9781450349253},
publisher = {Association for Computing Machinery},
address = {New York, NY, USA},
url = {https://doi.org/10.1145/3176349.3176384},
doi = {10.1145/3176349.3176384},
booktitle = {Proceedings of the 2018 Conference on Human Information Interaction \& Retrieval},
pages = {171–180},
numpages = {10},
location = {New Brunswick, NJ, USA},
series = {CHIIR '18}
}

@inproceedings{elSayed2024SAProcessMantra,
author = {Elsayed, Neven and Marriott, Kim and Smith, Ross and Thomas, Bruce H},
title = {Situated Analytics Process and Mantra},
year = {2024},
isbn = {9798400703317},
publisher = {Association for Computing Machinery},
address = {New York, NY, USA},
url = {https://doi.org/10.1145/3613905.3650814},
doi = {10.1145/3613905.3650814},
booktitle = {Extended Abstracts of the CHI Conference on Human Factors in Computing Systems},
articleno = {292},
numpages = {9},
location = {Honolulu, HI, USA},
series = {CHI EA '24}
}

@article{theNewStatistics2014,
author = {Geoff Cumming},
title ={The New Statistics: Why and How},
journal = {Psychological Science},
volume = {25},
number = {1},
pages = {7-29},
year = {2014},
doi = {10.1177/0956797613504966},
note ={PMID: 24220629},
URL = {https://doi.org/10.1177/0956797613504966},
eprint = {https://doi.org/10.1177/0956797613504966}
}

@incollection{dragicevic2016fair,
author={Dragicevic, Pierre},
editor={Robertson, Judy
and Kaptein, Maurits},
title={Fair Statistical Communication in HCI},
bookTitle={Modern Statistical Methods for HCI},
year={2016},
publisher={Springer International Publishing},
pages={291--330},
DOI={10.1007/978-3-319-26633-6_13},
HAL_ID = {hal-01377894},
}

@article{Hart2006NasaTlx,
author = {Sandra G. Hart},
title ={Nasa-Task Load Index (NASA-TLX); 20 Years Later},
journal = {Proceedings of the Human Factors and Ergonomics Society Annual Meeting},
volume = {50},
number = {9},
pages = {904-908},
year = {2006},
doi = {10.1177/154193120605000909},
URL = {https://doi.org/10.1177/154193120605000909},
eprint = {https://doi.org/10.1177/154193120605000909}
}

@inproceedings{lewis2013umux,
author = {Lewis, James R. and Utesch, Brian S. and Maher, Deborah E.},
title = {UMUX-LITE: When There's No Time for the SUS},
year = {2013},
isbn = {9781450318990},
publisher = {Association for Computing Machinery},
doi = {10.1145/2470654.2481287},
booktitle = {Proceedings of the SIGCHI Conference on Human Factors in Computing Systems},
pages = {2099–2102},
numpages = {4},
location = {Paris, France},
series = {CHI '13}
}

@article{IPQ2024,
author = {Tran, Tanh Quang and Langlotz, Tobias and Young, Jacob and Schubert, Thomas W. and Regenbrecht, Holger},
title = {Classifying Presence Scores: Insights and Analysis from Two Decades of the Igroup Presence Questionnaire (IPQ)},
year = {2024},
issue_date = {October 2024},
publisher = {Association for Computing Machinery},
address = {New York, NY, USA},
volume = {31},
number = {5},
issn = {1073-0516},
url = {https://doi.org/10.1145/3689046},
doi = {10.1145/3689046},
journal = {ACM Trans. Comput.-Hum. Interact.},
month = nov,
articleno = {61},
numpages = {26}
}

@inproceedings{Ens2021,
author = {Ens, Barrett and Bach, Benjamin and Cordeil, Maxime and Engelke, Ulrich and Serrano, Marcos and Willett, Wesley and Prouzeau, Arnaud and Anthes, Christoph and B\"{u}schel, Wolfgang and Dunne, Cody and Dwyer, Tim and Grubert, Jens and Haga, Jason H. and Kirshenbaum, Nurit and Kobayashi, Dylan and Lin, Tica and Olaosebikan, Monsurat and Pointecker, Fabian and Saffo, David and Saquib, Nazmus and Schmalstieg, Dieter and Szafir, Danielle Albers and Whitlock, Matt and Yang, Yalong},
title = {Grand Challenges in Immersive Analytics},
year = {2021},
doi = {10.1145/3411764.3446866},
booktitle = {Proceedings of the 2021 CHI Conference on Human Factors in Computing Systems},
articleno = {459},
location = {Yokohama, Japan},
series = {CHI '21}
}

@inproceedings{Sousa2023,
author = {Calepso, Aimee Sousa and Fleck, Philipp and Schmalstieg, Dieter and Sedlmair, Michael},
title = {Exploring Augmented Reality for Situated Analytics with Many Movable Physical Referents},
year = {2023},
isbn = {9798400703287},
publisher = {Association for Computing Machinery},
address = {New York, NY, USA},
url = {https://doi.org/10.1145/3611659.3615700},
doi = {10.1145/3611659.3615700},
booktitle = {Proceedings of the 29th ACM Symposium on Virtual Reality Software and Technology},
articleno = {6},
numpages = {12},
location = {Christchurch, New Zealand},
series = {VRST '23}
}

@article{Erickson2020,
	title = {A {Review} of {Visual} {Perception} {Research} in {Optical} {See}-{Through} {Augmented} {Reality}},
	doi = {10.2312/EGVE.20201256},
	journal = {Proceedings of the International Conference on Artificial Reality and Telexistence and Eurographics Symposium on Virtual Environments},
	author = {Erickson, A and Kim, K and Bruder, G and Welch, G},
	year = {2020},
	pages = {27--35},
}

@inproceedings{Bowman2003,
author = {Bowman, Doug A. and North, Chris and Chen, Jian and Polys, Nicholas F. and Pyla, Pardha S. and Yilmaz, Umur},
title = {Information-rich virtual environments: theory, tools, and research agenda},
year = {2003},
doi = {10.1145/1008653.1008669},
booktitle = {Proceedings of the ACM Symposium on Virtual Reality Software and Technology},
pages = {81–90},
location = {Osaka, Japan},
series = {VRST '03}
}

@inproceedings{Prouzeau2019VisualLink,
author = {Prouzeau, Arnaud and Lhuillier, Antoine and Ens, Barrett and Weiskopf, Daniel and Dwyer, Tim},
title = {Visual Link Routing in Immersive Visualisations},
year = {2019},
isbn = {9781450368919},
publisher = {Association for Computing Machinery},
address = {New York, NY, USA},
url = {https://doi.org/10.1145/3343055.3359709},
doi = {10.1145/3343055.3359709},
booktitle = {Proceedings of the 2019 ACM International Conference on Interactive Surfaces and Spaces},
pages = {241–253},
numpages = {13},
location = {Daejeon, Republic of Korea},
series = {ISS '19}
}

@ARTICLE{ChromaGazer2025Tosa,
  author={Tosa, Rinto and Hattori, Shingo and Hiroi, Yuichi and Itoh, Yuta and Hiraki, Takefumi},
  journal={IEEE Transactions on Visualization and Computer Graphics}, 
  title={ChromaGazer: Unobtrusive Visual Modulation Using Imperceptible Color Vibration for Visual Guidance}, 
  year={2025},
  volume={},
  number={},
  pages={1-9},
  doi={10.1109/TVCG.2025.3549173}}

@INPROCEEDINGS{satowski2018abovebelow,
  author={Satkowski, Marc and Rzayev, Rufat and Goebel, Eva and Dachselt, Raimund},
  booktitle={2022 IEEE International Symposium on Mixed and Augmented Reality (ISMAR)}, 
  title={ABOVE \& BELOW: Investigating Ceiling and Floor for Augmented Reality Content Placement}, 
  year={2022},
  volume={},
  number={},
  pages={518-527},
  doi={10.1109/ISMAR55827.2022.00068}}

@ARTICLE{Lee2024DesignPatterns,
  author={Lee, Benjamin and Sedlmair, Michael and Schmalstieg, Dieter},
  journal={IEEE Transactions on Visualization and Computer Graphics}, 
  title={Design Patterns for Situated Visualization in Augmented Reality}, 
  year={2024},
  volume={30},
  number={1},
  pages={1324-1335},
  doi={10.1109/TVCG.2023.3327398}}

@INPROCEEDINGS{quijano2024BrushandLink,
  author={Quijano-Chavez, Carlos and Doerr, Nina and Lee, Benjamin and Schmalstieg, Dieter and Sedlmair, Michael},
  booktitle={2024 IEEE Conference on Virtual Reality and 3D User Interfaces Abstracts and Workshops (VRW)}, 
  title={Brushing and Linking for Situated Analytics}, 
  year={2024},
  volume={},
  number={},
  pages={597-603},
  doi={10.1109/VRW62533.2024.00116}}

@article{shi2023Gazeselection,
author = {Shi, Rongkai and Wei, Yushi and Qin, Xueying and Hui, Pan and Liang, Hai-Ning},
title = {Exploring Gaze-assisted and Hand-based Region Selection in Augmented Reality},
year = {2023},
issue_date = {May 2023},
publisher = {Association for Computing Machinery},
address = {New York, NY, USA},
volume = {7},
number = {ETRA},
url = {https://doi.org/10.1145/3591129},
doi = {10.1145/3591129},
journal = {Proc. ACM Hum.-Comput. Interact.},
month = may,
articleno = {160},
numpages = {19}
}

@Inbook{Thomas2018,
author={Thomas, Bruce H.
and Welch, Gregory F.
and Dragicevic, Pierre
and Elmqvist, Niklas
and Irani, Pourang
and Jansen, Yvonne
and Schmalstieg, Dieter
and Tabard, Aur{\'e}lien
and ElSayed, Neven A. M.
and Smith, Ross T.
and Willett, Wesley},
title={Situated Analytics},
bookTitle={Immersive Analytics},
year={2018},
publisher={Springer International Publishing},
address={Cham},
pages={185--220},
isbn={978-3-030-01388-2},
doi={10.1007/978-3-030-01388-2_7},
url={https://doi.org/10.1007/978-3-030-01388-2_7}
}

@ARTICLE{sungbok2024SurveySita,
  author={Shin, Sungbok and Batch, Andrea and Butcher, Peter William Scott and Ritsos, Panagiotis D. and Elmqvist, Niklas},
  journal={IEEE Transactions on Visualization and Computer Graphics}, 
  title={The Reality of the Situation: A Survey of Situated Analytics}, 
  year={2024},
  volume={30},
  number={8},
  pages={5147-5164},
  doi={10.1109/TVCG.2023.3285546}}

@INPROCEEDINGS{elsayed2015situatedanalytics,
  author={ElSayed, Neven A. M. and Thomas, Bruce H. and Smith, Ross T. and Marriott, Kim and Piantadosi, Julia},
  booktitle={2015 IEEE Virtual Reality (VR)}, 
  title={Using augmented reality to support situated analytics}, 
  year={2015},
  volume={},
  number={},
  pages={175-176},
  doi={10.1109/VR.2015.7223352}}

@ARTICLE{lacoche2022arvrsimulation,
AUTHOR={Lacoche, J{\'e}r{\'e}my and Villain, Eric and Foulonneau, Anthony},
TITLE={Evaluating Usability and User Experience of AR Applications in VR Simulation},
JOURNAL={Frontiers in Virtual Reality},
VOLUME={3},
YEAR={2022},
URL={https://www.frontiersin.org/journals/virtual-reality/articles/10.3389/frvir.2022.881318},
DOI={10.3389/frvir.2022.881318},
ISSN={2673-4192}}

@ARTICLE{fleck2023ragrug,
  author={Fleck, Philipp and Calepso, Aimée Sousa and Hubenschmid, Sebastian and Sedlmair, Michael and Schmalstieg, Dieter},
  journal={IEEE Transactions on Visualization and Computer Graphics}, 
  title={RagRug: A Toolkit for Situated Analytics}, 
  year={2023},
  volume={29},
  number={7},
  pages={3281-3297},
  doi={10.1109/TVCG.2022.3157058}}

@ARTICLE{yu2024persival,
  author={Yu, Xingyao and Rosin, David and Kässinger, Johannes and Lee, Benjamin and Dürr, Frank and Becker, Christian and Röhrle, Oliver and Sedlmair, Michael},
  journal={IEEE Computer Graphics and Applications}, 
  title={PerSiVal: On-Body AR Visualization of Biomechanical Arm Simulations}, 
  year={2024},
  volume={44},
  number={6},
  pages={24-38},
  doi={10.1109/MCG.2024.3494598}}

@inproceedings{luo2023pearl,
   author = {Weizhou Luo and Zhongyuan Yu and Rufat Rzayev and Marc Satkowski and Stefan Gumhold and Matthew McGinity and Raimund Dachselt},
   title = {Pearl: Physical Environment based Augmented Reality Lenses for In-Situ Human Movement Analysis},
   booktitle = {Proceedings of the 2023 CHI Conference on Human Factors in Computing Systems},
   series = {CHI '23},
   number = {381},
   year = {2023},
   month = {04},
   isbn = {9781450394215},
   location = {Hamburg, Germany},
   numpages = {15},
   doi = {10.1145/3544548.3580715},
   url = {https://doi.org/10.1145/3544548.3580715},
   publisher = {Association for Computing Machinery},
   address = {New York, NY, USA}
}

@inproceedings{lin2021basket,
author = {Lin, Tica and Singh, Rishi and Yang, Yalong and Nobre, Carolina and Beyer, Johanna and Smith, Maurice A. and Pfister, Hanspeter},
title = {Towards an Understanding of Situated AR Visualization for Basketball Free-Throw Training},
year = {2021},
isbn = {9781450380966},
publisher = {Association for Computing Machinery},
address = {New York, NY, USA},
url = {https://doi.org/10.1145/3411764.3445649},
doi = {10.1145/3411764.3445649},
booktitle = {Proceedings of the 2021 CHI Conference on Human Factors in Computing Systems},
articleno = {461},
numpages = {13},
location = {Yokohama, Japan},
series = {CHI '21}
}

@inproceedings{ye2023Proobjar,
author = {Ye, Hui and Leng, Jiaye and Xiao, Chufeng and Wang, Lili and Fu, Hongbo},
title = {ProObjAR: Prototyping Spatially-aware Interactions of Smart Objects with AR-HMD},
year = {2023},
isbn = {9781450394215},
publisher = {Association for Computing Machinery},
address = {New York, NY, USA},
url = {https://doi.org/10.1145/3544548.3580750},
doi = {10.1145/3544548.3580750},
booktitle = {Proceedings of the 2023 CHI Conference on Human Factors in Computing Systems},
articleno = {457},
numpages = {15},
location = {Hamburg, Germany},
series = {CHI '23}
}

@inproceedings{tong2019actionUnits,
author = {Tong, Lingwei and Jung, Sungchul and Lindeman, Robert W.},
title = {Action Units: Directing User Attention in 360-degree Video based VR},
year = {2019},
isbn = {9781450370011},
publisher = {Association for Computing Machinery},
address = {New York, NY, USA},
url = {https://doi.org/10.1145/3359996.3364706},
doi = {10.1145/3359996.3364706},
booktitle = {Proceedings of the 25th ACM Symposium on Virtual Reality Software and Technology},
articleno = {52},
numpages = {2},
location = {Parramatta, NSW, Australia},
series = {VRST '19}
}

@inproceedings{baldonado2000cmv,
author = {Wang Baldonado, Michelle Q. and Woodruff, Allison and Kuchinsky, Allan},
title = {Guidelines for using multiple views in information visualization},
year = {2000},
isbn = {1581132522},
publisher = {Association for Computing Machinery},
address = {New York, NY, USA},
url = {https://doi.org/10.1145/345513.345271},
doi = {10.1145/345513.345271},
booktitle = {Proceedings of the Working Conference on Advanced Visual Interfaces},
pages = {110–119},
numpages = {10},
location = {Palermo, Italy},
series = {AVI '00}
}

@inproceedings{kim2025pinchcatcher,
author = {Kim, Jinwook and Park, Sangmin and Zhou, Qiushi and Gonzalez-Franco, Mar and Lee, Jeongmi and Pfeuffer, Ken},
title = {PinchCatcher: Enabling Multi-selection for Gaze+Pinch},
year = {2025},
isbn = {9798400713941},
publisher = {Association for Computing Machinery},
address = {New York, NY, USA},
url = {https://doi.org/10.1145/3706598.3713530},
doi = {10.1145/3706598.3713530},
booktitle = {Proceedings of the 2025 CHI Conference on Human Factors in Computing Systems},
articleno = {853},
numpages = {16},
location = {
},
series = {CHI '25}
}

@inproceedings{Wu2023AcquisitionOclussion,
author = {Wu, Zhiqing and Yu, Difeng and Goncalves, Jorge},
title = {Point- and Volume-Based Multi-object Acquisition in VR},
year = {2023},
isbn = {978-3-031-42279-9},
publisher = {Springer-Verlag},
address = {Berlin, Heidelberg},
url = {https://doi.org/10.1007/978-3-031-42280-5_2},
doi = {10.1007/978-3-031-42280-5_2},
booktitle = {Human-Computer Interaction – INTERACT 2023: 19th IFIP TC13 International Conference, York, UK, August 28 – September 1, 2023, Proceedings, Part I},
pages = {20–42},
numpages = {23},
location = {York, United Kingdom}
}

@ARTICLE{Koytek2018MyBrush,
  author={Koytek, Philipp and Perin, Charles and Vermeulen, Jo and André, Elisabeth and Carpendale, Sheelagh},
  journal={IEEE Transactions on Visualization and Computer Graphics}, 
  title={MyBrush: Brushing and Linking with Personal Agency}, 
  year={2018},
  volume={24},
  number={1},
  pages={605-615},
  doi={10.1109/TVCG.2017.2743859}}

@INPROCEEDINGS{mahmood2018multipleviewsAR,
  author={Mahmood, Tahir and Butler, Erik and Davis, Nicholas and Huang, Jian and Lu, Aidong},
  booktitle={2018 International Symposium on Big Data Visual and Immersive Analytics (BDVA)}, 
  title={Building Multiple Coordinated Spaces for Effective Immersive Analytics through Distributed Cognition}, 
  year={2018},
  volume={},
  number={},
  pages={1-11},
  doi={10.1109/BDVA.2018.8533893}}

@inproceedings{Hubenschmid2021STREAM,
author = {Hubenschmid, Sebastian and Zagermann, Johannes and Butscher, Simon and Reiterer, Harald},
title = {STREAM: Exploring the Combination of Spatially-Aware Tablets with Augmented Reality Head-Mounted Displays for Immersive Analytics},
year = {2021},
isbn = {9781450380966},
publisher = {Association for Computing Machinery},
address = {New York, NY, USA},
url = {https://doi.org/10.1145/3411764.3445298},
doi = {10.1145/3411764.3445298},
booktitle = {Proceedings of the 2021 CHI Conference on Human Factors in Computing Systems},
articleno = {469},
numpages = {14},
location = {Yokohama, Japan},
series = {CHI '21}
}

@inproceedings{langner2021marvis,
author = {Langner, Ricardo and Satkowski, Marc and B\"{u}schel, Wolfgang and Dachselt, Raimund},
title = {MARVIS: Combining Mobile Devices and Augmented Reality for Visual Data Analysis},
year = {2021},
isbn = {9781450380966},
publisher = {Association for Computing Machinery},
address = {New York, NY, USA},
url = {https://doi.org/10.1145/3411764.3445593},
doi = {10.1145/3411764.3445593},
booktitle = {Proceedings of the 2021 CHI Conference on Human Factors in Computing Systems},
articleno = {468},
numpages = {17},
location = {Yokohama, Japan},
series = {CHI '21}
}

@article{surveyXVR2022,
author = {Fröhler, B. and Anthes, C. and Pointecker, F. and Friedl, J. and Schwajda, D. and Riegler, A. and Tripathi, S. and Holzmann, C. and Brunner, M. and Jodlbauer, H. and Jetter, H.-C. and Heinzl, C.},
title = {A Survey on Cross-Virtuality Analytics},
journal = {Computer Graphics Forum},
volume = {41},
number = {1},
pages = {465-494},
doi = {https://doi.org/10.1111/cgf.14447},
url = {https://onlinelibrary.wiley.com/doi/abs/10.1111/cgf.14447},
eprint = {https://onlinelibrary.wiley.com/doi/pdf/10.1111/cgf.14447},
year = {2022}
}

@ARTICLE{Borowski2025dashspace,
  author={Borowski, Marcel and Butcher, Peter W. S. and Kristensen, Janus Bager and Petersen, Jonas Oxenbøll and Ritsos, Panagiotis D. and Klokmose, Clemens N. and Elmqvist, Niklas},
  journal={IEEE Transactions on Visualization and Computer Graphics}, 
  title={DashSpace: A Live Collaborative Platform for Immersive and Ubiquitous Analytics}, 
  year={2025},
  volume={},
  number={},
  pages={1-13},
  doi={10.1109/TVCG.2025.3537679}}

@inproceedings{satriadi2023Proxsituated,
author = {Satriadi, Kadek Ananta and Cunningham, Andrew and Smith, Ross T. and Dwyer, Tim and Drogemuller, Adam and Thomas, Bruce H.},
title = {ProxSituated Visualization: An Extended Model of Situated Visualization using Proxies for Physical Referents},
year = {2023},
isbn = {9781450394215},
publisher = {Association for Computing Machinery},
address = {New York, NY, USA},
url = {https://doi.org/10.1145/3544548.3580952},
doi = {10.1145/3544548.3580952},
booktitle = {Proceedings of the 2023 CHI Conference on Human Factors in Computing Systems},
articleno = {382},
numpages = {20},
location = {Hamburg, Germany},
series = {CHI '23}
}

@inproceedings{wallgrun2019vrexperience,
author = {Wallgr\"{u}n, Jan Oliver and Chang, Jack (Shen-Kuen) and Zhao, Jiayan and Sajjadi, Pejman and Oprean, Danielle and Murphy, Thomas B. and Baka, Jennifer and Klippel, Alexander},
title = {For the Many, Not the One: Designing Low-Cost Joint VR Experiences for Place-Based Learning},
year = {2019},
isbn = {978-3-030-31907-6},
publisher = {Springer-Verlag},
address = {Berlin, Heidelberg},
url = {https://doi.org/10.1007/978-3-030-31908-3_9},
doi = {10.1007/978-3-030-31908-3_9},
booktitle = {Virtual Reality and Augmented Reality: 16th EuroVR International Conference, EuroVR 2019, Tallinn, Estonia, October 23–25, 2019, Proceedings},
pages = {126–148},
numpages = {23},
location = {Tallinn, Estonia}
}

@inproceedings{Sidenmark2020selectionVR,
author = {Sidenmark, Ludwig and Clarke, Christopher and Zhang, Xuesong and Phu, Jenny and Gellersen, Hans},
title = {Outline Pursuits: Gaze-assisted Selection of Occluded Objects in Virtual Reality},
year = {2020},
isbn = {9781450367080},
publisher = {Association for Computing Machinery},
address = {New York, NY, USA},
url = {https://doi.org/10.1145/3313831.3376438},
doi = {10.1145/3313831.3376438},
booktitle = {Proceedings of the 2020 CHI Conference on Human Factors in Computing Systems},
pages = {1–13},
numpages = {13},
location = {Honolulu, HI, USA},
series = {CHI '20}
}

@article{grogorick2018vgmethods,
author = {Grogorick, Steve and Albuquerque, Georgia and Tauscher, Jan-Philipp and Magnor, Marcus},
title = {Comparison of Unobtrusive Visual Guidance Methods in an Immersive Dome Environment},
year = {2018},
issue_date = {October 2018},
publisher = {Association for Computing Machinery},
address = {New York, NY, USA},
volume = {15},
number = {4},
issn = {1544-3558},
url = {https://doi.org/10.1145/3238303},
doi = {10.1145/3238303},
journal = {ACM Trans. Appl. Percept.},
month = sep,
articleno = {27},
numpages = {11}
}

@inproceedings{lin2017assistingvideos,
author = {Lin, Yen-Chen and Chang, Yung-Ju and Hu, Hou-Ning and Cheng, Hsien-Tzu and Huang, Chi-Wen and Sun, Min},
title = {Tell Me Where to Look: Investigating Ways for Assisting Focus in 360° Video},
year = {2017},
isbn = {9781450346559},
publisher = {Association for Computing Machinery},
address = {New York, NY, USA},
url = {https://doi.org/10.1145/3025453.3025757},
doi = {10.1145/3025453.3025757},
booktitle = {Proceedings of the 2017 CHI Conference on Human Factors in Computing Systems},
pages = {2535–2545},
numpages = {11},
location = {Denver, Colorado, USA},
series = {CHI '17}
}

@inproceedings{Gruenefeld2018haloswedge,
author = {Gruenefeld, Uwe and Ali, Abdallah El and Boll, Susanne and Heuten, Wilko},
title = {Beyond Halo and Wedge: visualizing out-of-view objects on head-mounted virtual and augmented reality devices},
year = {2018},
isbn = {9781450358989},
publisher = {Association for Computing Machinery},
address = {New York, NY, USA},
url = {https://doi.org/10.1145/3229434.3229438},
doi = {10.1145/3229434.3229438},
booktitle = {Proceedings of the 20th International Conference on Human-Computer Interaction with Mobile Devices and Services},
articleno = {40},
numpages = {11},
location = {Barcelona, Spain},
series = {MobileHCI '18}
}

@article{Fox2023VisualCuesLowVision,
author = {Dylan R. Fox and Ahmad Ahmadzada and Clara T. Friedman and Shiri Azenkot and Marlena A. Chu and Roberto Manduchi and Emily A. Cooper},
journal = {Opt. Express},
number = {4},
pages = {6827--6848},
publisher = {Optica Publishing Group},
title = {Using augmented reality to cue obstacles for people with low vision},
volume = {31},
month = {Feb},
year = {2023},
url = {https://opg.optica.org/oe/abstract.cfm?URI=oe-31-4-6827},
doi = {10.1364/OE.479258},
}

@INPROCEEDINGS{Gutkowski2021tourvideoAR,
  author={Gutkowski, Nicolas and Quigley, Paul and Ogle, Todd and Hicks, David and Taylor, Jessica and Tucker, Thomas and Bowman, Doug A.},
  booktitle={2021 IEEE International Symposium on Mixed and Augmented Reality Adjunct (ISMAR-Adjunct)}, 
  title={Designing Historical Tours for Head-Worn AR}, 
  year={2021},
  volume={},
  number={},
  pages={26-33},
  doi={10.1109/ISMAR-Adjunct54149.2021.00016}}

@inproceedings{sutton2022saliencyVG,
author = {Sutton, Jonathan and Langlotz, Tobias and Plopski, Alexander and Zollmann, Stefanie and Itoh, Yuta and Regenbrecht, Holger},
title = {Look over there! Investigating Saliency Modulation for Visual Guidance with Augmented Reality Glasses},
year = {2022},
isbn = {9781450393201},
publisher = {Association for Computing Machinery},
address = {New York, NY, USA},
url = {https://doi.org/10.1145/3526113.3545633},
doi = {10.1145/3526113.3545633},
booktitle = {Proceedings of the 35th Annual ACM Symposium on User Interface Software and Technology},
articleno = {81},
numpages = {15},
location = {Bend, OR, USA},
series = {UIST '22}
}

@INPROCEEDINGS{Yokomi2021VGBrightness,
  author={Yokomi, Masatoshi and Isoyama, Naoya and Sakata, Nobuchika and Kiyokawa, Kiyoshi},
  booktitle={2021 IEEE Conference on Virtual Reality and 3D User Interfaces Abstracts and Workshops (VRW)}, 
  title={Subtle Gaze Guidance for 360° Content by Gradual Brightness Modulation and Termination of Modulation by Gaze Approaching}, 
  year={2021},
  volume={},
  number={},
  pages={520-521},
  doi={10.1109/VRW52623.2021.00142}}

@Article{Krzywinski2013pvalues,
author={Krzywinski, Martin
and Altman, Naomi},
title={Error bars},
journal={Nature Methods},
year={2013},
month={Oct},
day={01},
volume={10},
number={10},
pages={921-922},
issn={1548-7105},
doi={10.1038/nmeth.2659},
url={https://doi.org/10.1038/nmeth.2659}
}

@article{Lei2023ARVisualSalience,
author = {Xin Lei and Yueh-Lin Tsai and Pei-Luen Patrick Rau and},
title = {Harnessing the Visual Salience Effect With Augmented Reality to Enhance Relevant Information and to Impair Distracting Information},
journal = {International Journal of Human–Computer Interaction},
volume = {39},
number = {6},
pages = {1280--1293},
year = {2023},
publisher = {Taylor \& Francis},
doi = {10.1080/10447318.2022.2062548},
URL = {https://doi.org/10.1080/10447318.2022.2062548
},
eprint = {https://doi.org/10.1080/10447318.2022.2062548
}

}

@ARTICLE{Assor2024NonVisibleSita,
  author={Assor, Ambre and Prouzeau, Arnaud and Hachet, Martin and Dragicevic, Pierre},
  journal={IEEE Transactions on Visualization and Computer Graphics}, 
  title={Handling Non-Visible Referents in Situated Visualizations}, 
  year={2024},
  volume={30},
  number={1},
  pages={1336-1346},
  doi={10.1109/TVCG.2023.3327361}}

@inproceedings{Buja1991Interactive,
	location = {Washington, {DC}, {USA}},
	title = {Interactive data visualization using focusing and linking},
	isbn = {978-0-8186-2245-8},
	series = {{VIS} '91},
	pages = {156--163},
	booktitle = {Proceedings of the 2nd conference on Visualization '91},
	publisher = {{IEEE} Computer Society Press},
	author = {Buja, Andreas and {McDonald}, John Alan and Michalak, John and Stuetzle, Werner},
	urldate = {2025-06-25},
	date = {1991-10-22},
    year = {1991},
    doi = {10.1109/VISUAL.1991.175794}
}

@inproceedings{Grandi2021Design,
	title = {Design and Simulation of Next-Generation Augmented Reality User Interfaces in Virtual Reality},
	url = {https://ieeexplore.ieee.org/document/9419218},
	doi = {10.1109/VRW52623.2021.00011},
	eventtitle = {2021 {IEEE} Conference on Virtual Reality and 3D User Interfaces Abstracts and Workshops ({VRW})},
	pages = {23--29},
	booktitle = {2021 {IEEE} Conference on Virtual Reality and 3D User Interfaces Abstracts and Workshops ({VRW})},
	author = {Grandi, Jerônimo G and Cao, Zekun and Ogren, Mark and Kopper, Regis},
	urldate = {2025-06-25},
	date = {2021-03},
    year = {2021}
}

@article{Lee2013Effects,
	title = {The Effects of Visual Realism on Search Tasks in Mixed Reality Simulation},
	volume = {19},
	issn = {1941-0506},
	url = {https://ieeexplore.ieee.org/document/6479181},
	doi = {10.1109/TVCG.2013.41},
	pages = {547--556},
	number = {4},
    journal={IEEE Transactions on Visualization and Computer Graphics}, 
	journaltitle = {{IEEE} Transactions on Visualization and Computer Graphics},
	author = {Lee, Cha and Rincon, Gustavo A. and Meyer, Greg and Höllerer, Tobias and Bowman, Doug A.},
	urldate = {2025-06-25},
	date = {2013-04},
    year = {2013}
}

@inproceedings{syiem2021attentionTunnelingAR,
author = {Syiem, Brandon Victor and Kelly, Ryan M. and Goncalves, Jorge and Velloso, Eduardo and Dingler, Tilman},
title = {Impact of Task on Attentional Tunneling in Handheld Augmented Reality},
year = {2021},
isbn = {9781450380966},
publisher = {Association for Computing Machinery},
address = {New York, NY, USA},
url = {https://doi.org/10.1145/3411764.3445580},
doi = {10.1145/3411764.3445580},
booktitle = {Proceedings of the 2021 CHI Conference on Human Factors in Computing Systems},
articleno = {193},
numpages = {14},
location = {Yokohama, Japan},
series = {CHI '21}
}

@inproceedings{Sutton2024FlickerVisualGuidance,
author = {Sutton, Jonathan and Langlotz, Tobias and Plopski, Alexander and Hornb\ae{}k, Kasper},
title = {Flicker Augmentations: Rapid Brightness Modulation for Real-World Visual Guidance using Augmented Reality},
year = {2024},
isbn = {9798400703300},
publisher = {Association for Computing Machinery},
address = {New York, NY, USA},
url = {https://doi.org/10.1145/3613904.3642085},
doi = {10.1145/3613904.3642085},
booktitle = {Proceedings of the 2024 CHI Conference on Human Factors in Computing Systems},
articleno = {752},
numpages = {19},
location = {Honolulu, HI, USA},
series = {CHI '24}
}

@ARTICLE{Jung2021RealismPresenceVR,
AUTHOR={Jung, Sungchul  and Lindeman, Robert. W },     
TITLE={Perspective: Does Realism Improve Presence in VR? Suggesting a Model and Metric for VR Experience Evaluation},     
JOURNAL={Frontiers in Virtual Reality},     
VOLUME={Volume 2 - 2021},
YEAR={2021},
URL={https://www.frontiersin.org/journals/virtual-reality/articles/10.3389/frvir.2021.693327},
DOI={10.3389/frvir.2021.693327},
ISSN={2673-4192},
}

@ARTICLE{Willet2017EmbeddedData,
  author={Willett, Wesley and Jansen, Yvonne and Dragicevic, Pierre},
  journal={IEEE Transactions on Visualization and Computer Graphics}, 
  title={Embedded Data Representations}, 
  year={2017},
  volume={23},
  number={1},
  pages={461-470},
  doi={10.1109/TVCG.2016.2598608}}

\end{document}